\documentclass[a4paper]{jpconf}
\usepackage{graphicx}
\usepackage{bm}
\def\be{\begin{equation}}
\def\ee{\end{equation}}
\def\br{\begin{eqnarray}}
\def\er{\end{eqnarray}}
\def\brn{\begin{eqnarray*}}
\def\ern{\end{eqnarray*}}

\def\bit{\begin{itemize}}
\def\eit{\end{itemize}}
\def\bnu{\begin{enumerate}}
\def\enu{\end{enumerate}}
\def\ie{{\em i.e., }}
\def\rf#1{{(\ref{#1})}}

\def\etal{{\it et al.}}

\newcommand{\Mass}{\mathrm{M}}

\def\F {{{\cal F}}}

\begin{document}
\title{Nonmesonic Hyperon Weak Decay Spectra in ${}^{12}_{\Lambda}$C}
\author{I. Gonzalez$^{1,2}$, A. Deppman$^{1}$,  S. Duarte$^{3}$, F. Krmpoti\'c$^{4}$,
 M. S. Hussein$^{1}$, C. Barbero$^{4,5}$}
\address{$^1$ Instituto de F\'isica, Universidad de S\~ao Paulo, S\~ao Paulo, Brasil}
\address{$^2$ Instituto de Tecnolog\'ias y Ciencias Aplicadas, Havana, Cuba}
\address{$^3$ Centro Brasileiro de Pesquisas F\'isicas, Rio de Janeiro, Brasil}
\address{$^4$ Instituto de F\'isica La Plata, Universidad Nacional de La Plata, La Plata, Argentina}
\address{$^5$ Departamento de F\'{\i}sica, Universidad Nacional de
La Plata, La Plata, Argentina}

\ead{israel.gonzalezmedina@gmail.com, adeppman@gmail.com, sbd@cbpf.br,
krmpotic@fisica.unlp.edu.ar, hussein@if.usp.br,
barbero@fisica.unlp.edu.ar}

\begin{abstract}
We study the nonmesonic weak decay (NMWD) $\Lambda N \to nN$ of
the ${}^{12}_{\Lambda} C$ hypernucleus induced by the nucleon
$N=n,p$ with  transition rate  $\Gamma_N$. The  nuclear process is
described by the interplay of two models; one describing the NMWD
of hyperon $\Lambda$  in the nuclear environment, and the other
taking into account the Final State Interaction (FSI) of the two
outgoing nucleons with the residual nucleus. The first one is done
in the framework of the Independent-Particle Shell-Model (IPSM),
with the decay dynamics represented by the exchange of
 $\pi+ \eta+ K+\rho+\omega+K^*$
  mesons with usual parametrization. For the second one is used
a time dependent multicollisional intranuclear cascade  schema
(implemented in the CRISP code - Collaboration Rio-S\~ao Paulo).
The results obtained for inclusive and exclusive kinetic energy
spectra, and the angular correlation are compared with recent data
from KEK and FINUDA experiments.  The calculated ratio $(\Gamma_n / \Gamma_p)^{\mbox{\tiny FSI}}$,
between the numbers of emitted back-to-back $nn$ and
$np$ pairs, is  in good agreement with the experimental data.
\end{abstract}

\section{Introduction}
 Since the mesonic channel $\Lambda \rightarrow \pi N$ is
strongly inhibited inside nuclei due to Pauli blocking, the
nonmesonic  weak decay (NMWD) $\Lambda N\rightarrow n N$
emerges as the dominant channel. This primary decay can be induced
by a neutron ($\Lambda n\rightarrow nn$) or a proton ($\Lambda
p\rightarrow np$) with widths $\Gamma_n$ and $\Gamma_p$,
respectively. During the last two decades quite significant
efforts have been invested in  solving the so called 'NMWD puzzle'.
 This puzzle is
related to the large discrepancy between earlier experimental data
and theoretical predictions on the NMWD  spectra, that yield information
about both the primary nonmesonic decay, and the subsequent Final
State Interactions (FSI)
  originated by the nuclear medium.
The main aim of the present work is to  analyze in  which way the
FSI  modify the spectra obtained within the Independent Particle
Shell Model (IPSM), and to find out to what extent it is possible
in this way to explain the experimental data.
The spectra that we are interested in  are: 1) the  single-nucleon spectra $S_{N}(E_N)$,
 as a function of one-nucleon energy $E_N$,
and 2) the two-particle-coincidence spectra as  a function of:
i) the
sum of  kinetic energies $E_n+E_N\equiv E_{nN}$, $S_{N}(E_{nN})$, and ii) the opening angle
$\theta_{nN}$, $S_{N}(\cos\theta_{nN})$.
The explicit relationships for these  transition probability densities  can be obtained
by performing derivatives on the appropriate equation for
$\Gamma_N$~\cite{Bar08,Bau09,Krm10,Krm10a}, \ie
\be
S_N(E_N)=\frac{d\Gamma_N}{dE_N},\hspace{0.3cm}S_N(E_{nN})=\frac{d\Gamma_N}{dE_{nN}}
,\hspace{0.3cm}S_N(\cos\theta_{nN})=\frac{d\Gamma_N}{d\cos\theta_{nN}}.
\label{1}\ee
The measured spectra are obtained by counting the number of
emitted nucleons $\Delta{\rm N}_N$ within
the energy  bin $\Delta{E}=10$ MeV, or the angular bin  $\Delta\cos\theta=0.05$, always
 corrected by the detection efficiency.  Here we  take advantage of the
fact that  in most of the KEK
experiments~\cite{Kim06,Kim09, Bha11} $\Delta{\rm N}_N$ are
 normalized to the number of NMWD processes, ${\rm N}_{NM}$, while for  the
FINUDA proton  spectra \cite{Agn10,Gar10}
we have at our disposal also the number of produced hypernuclei ${\rm N}_{H}$.
Therefore, we can explore the following relationships
  \be
\frac{\Delta{\rm N}_N}{{\rm N}_{NM}}=\frac{\Delta{\Gamma}_N}{{\Gamma}_{NM}},
\hspace{0.5cm}\frac{\Delta{\rm N}_N}{{\rm N}_{H}}=\frac{\Delta{\Gamma}_N}{{\Gamma}_{H}},
\label{2}\ee
where  $\Delta\Gamma_N$ is the emission rate of protons
within the experimentally
fixed bin, and $\Gamma_{NM}$ and $\Gamma_{H}$ are, respectively, the NMWD rate and
the total hypernuclear decay rate. It could be worth noting  that
$\Gamma_{H}\equiv \Gamma_{M}+\Gamma_{NM}$, \ie is the sum of mesonic  and nonmesonic
 decay rates. It is not clear whether one should use experimental or theoretical values for
 the rates $\Gamma_{NM}$, and $\Gamma_{H}$, and we will adopt the first possibility.
 Then the correspondences between the theory and data are
\br
\Delta\Gamma_N^{th}&\Longleftrightarrow &\Delta\Gamma_N^{exp}
=\left.{\Gamma}_{NM}
\frac{\Delta{\rm N}_N}{{\rm N}_{NM}}\right|_{KEK}
={\Gamma}_{H}
\left.\frac{\Delta{\rm N}_N}{{\rm N}_{H}}\right|_{FINUDA}.
\label{3}\er
For $_\Lambda^{12}$C ${\Gamma}_{NM}=0.95\pm0.04$~\cite{Kim09}, and
${\Gamma}_{H}=1.21\pm0.21$~\cite{Agn09}. The theoretical decay widths
$\Delta\Gamma_N^{th}$ are:
$\Delta\Gamma_N(E_N)=S_N(E_N)\Delta E$,
$\Delta\Gamma_N(E_{nN})=S_N(E_{nN})\Delta E$, and
$\Delta\Gamma_N(\cos\theta_{nN})=S_N(\cos\theta_{nN})\Delta \cos\theta$ for the spectra
explained in 1) and 2).  As the one-proton (one-neutron) induced decay prompts
the emission of an $np$ ($nn$) pair the total neutron kinetic energy width is
$\Delta\Gamma_{nt}(E_n)=(S_p(E_n)+2S_n(E_n))\Delta E$.
For $\Delta\Gamma_p(E_p)$,
 $\Delta\Gamma_{nt}(E_n)$, and
$\Delta\Gamma_N(\cos\theta_{nN})$ are  available the data, with the corresponding
errors.
 As for the
kinetic energy sums, only the
data for yields
 $\Delta Y_N(E_{nN})$  were reported so far, without specifying their
errors~\cite{Kim06}.
It is  not known how  $\Delta \Gamma_N(E_{nN})$ and $\Delta Y_N(E_{nN})$
 are related with each other,
and thus one is forced here to normalize
the theoretical results to the data in a similar way as done in Ref.~\cite{Bau09}, \ie
\be
{\Delta Y}_{N}(E_{nN})=
\frac{{\bar Y}_{N}^{exp}}{\bar{{{\Gamma}}}_N}{\bar S}_{N}(E_{nN})\Delta E,
\label{4}\ee
where
 \be
{\bar Y}_N^{exp}=\sum_{i=1}^{m} \Delta Y_N^{exp}(E^i_{nN}),\hspace{0.3cm}
{\bar{\Gamma}_N}=\sum_{i=1}^{m} S_{N}(E^i_{nN})\Delta E,
\label{5}\ee
with  $m$ being the number of kinetic energy sum bins, and
the barred  quantities  indicate that
and they are constructed with events constrained to $E_{N}> 30$  MeV,  and $\cos\theta_{nN}< -0.7$.

\section{Independent Particle Shell Model for the  primary  NMWD}
Within the IPSM  the spectra defined in \rf{1} read~\cite{Bar08,Bau09,Krm10,Krm10a}:
\br
S_N(E_{N})&=&
(A-2)\frac{8\Mass^3}{\pi}\sum_{j_N} \int _{-1}^{+1}
d\cos\theta_{nN}\sqrt{\frac{E_{N}}{E_N'}}\, E_n\, \F_{j_N}(pP),
\label{6}\er
\br
S_N(\cos\theta_{nN}) &=&
(A-2)\frac{8\Mass^3}{\pi}\sum_{j_N}
\int_0^{{\tilde E}_{j_N}} dE_{N}\sqrt{\frac{E_{N}}{E_N'}}\, E_n \F_{j_N}(pP),
\label{7}\er
\br
S_N(E_{nN}) &=&
\frac{4\Mass^3}{\pi}\sqrt{A(A-2)^3}\sum_{j_N}
\sqrt{(\Delta_{j_N}-E_{nN})(E_{nN}-{\tilde\Delta}_{j_N})} \F_{j_N}(pP),
\label{8}\er
where the summation goes over all occupied single-particle states $j\equiv \{nlj\}$
 for a given $N=p,n$, and
 \br
E'_N&=&(A-2)(A-1)\Delta_{j_N}-E_{N} [(A-1)^2-\cos^2\theta_{nN}],
\label{9}\er
  \br
  E_n&=&\left[\sqrt{E'_N} -\sqrt{E_{N}}\cos{\theta_{nN}}\right]^2(A-1)^{-2},
~~~{\tilde E}_{j_N}=\frac{A-1}{A}\Delta_{j_N},
\label{10}\er
\be
\Delta_{j_N} =\Delta  + \varepsilon_{\Lambda} +
\varepsilon_{j_N},~~~~{\tilde\Delta}_{j_N}=\Delta_{j_N}\frac{A-2}{A}.
\label{11}\ee
$A$ is the nuclear mass,  $\varepsilon$'s are the single-particle energies,
and $\Delta=\Mass_\Lambda-\Mass$ is the mass difference between the $\Lambda$-hyperon
and the nucleon.
The quantity  $\F_{j_N}(pP)$ contains the isospin dependence, as well as the dynamics,
which is described by exchanges of
complete pseudoscalar ($\pi, K, \eta$) and vector
($\rho,\omega,K^*$) meson octets with the
weak coupling constants from Refs.~\cite{Par97,Par01}.
It is defined in Eq. (42) of Ref.~\cite{Krm10}, and
depends on center of mass and relative momenta of the $nN$ pairs,
 \br
{P}&=&\sqrt{(A-2)(2\Mass\Delta_{j_N}- p_n^2 -p_N^2)},
~~~~p=\sqrt{\Mass \Delta_{j_N}- \frac{A}{4(A-2)} P^2},
 \label{12}\er
 with $p_n$, and  $p_N$
being the momenta of the emitted particles.

\section{Monte Carlo calculation of FSI with CRISP}

\begin{figure}[htpb]
\includegraphics[width=8.6cm,height=6.cm]{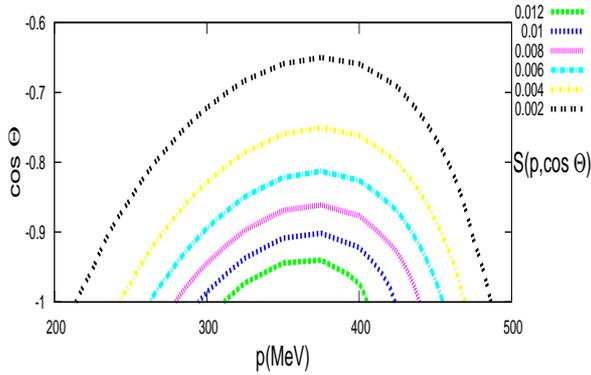}
\caption{Two-dimensional spectrum for protons $S(p_p,\cos\theta_{np})$
as evaluated from the IPSM  Eq. \rf{13}.}
 \label{Fig1}
\end{figure}

After having modelled the primary process, we investigate the FSI through
the time evolution of Monte Carlo samples constructed to represent the
decaying nuclear system, making use of
a time dependent multicollisional intranuclear cascade  schema
(implemented in the CRISP code - Collaboration Rio-S\~ao Paulo).
The initial sample configurations are spherically shaped
portions of cold Fermi gas
with $A-2$ nucleons, where are incorporated the two energetic intruder nucleons
resulting from the primary process, with the IPSM spectra
\br
S_N(p_N,\cos\theta_{nN})=\frac{d^2\Gamma_N}{dp_Nd\cos\theta_{nN}}=
(A-2)\frac{4\Mass}{\pi}\sum_{j_N}p_N^2 p_n^2\F_{j_N}(pP),
 \label{13}\er
which are shown in Fig. \ref{Fig1} for the emission of the $np$ pair,
and are the starting configuration for  the time evolution of the $A$-nucleon system,
 following the sequence of all possible binaries collisions between nucleons.
 A square well potential is included to mimic the nuclear surface,
 keeping  the low energy nucleons bound within  the system.
The depth of this potential well is properly chosen in order to have an unperturbed initial
 configuration without nucleon loss.
 In fact, the potential confines low energy nucleons through  internal reflections, and
 permits refractions/deflections
 of energetic nucleons,
   as dictated by the tunnelling probability of the charged particles through the nuclear
  Coulomb barrier~\cite{Dep02}.

The time order of the nucleon-nucleon collisions during
 the evolution of the system is reconstructed after each collision between pairs  of particles,
  and their displacement inside the nucleus.
  Consequently, the many-body configurations are followed step by step in phase space, and in time.
 The cascade  phase is ended when no particle is able to  leave the nuclear volume,
 and the number of bound nucleons
  in the residual nucleus does not change anymore.
  However, the system still can present some excitation energy compared to a cold nucleon Fermi gas.
  A thermal decay chain is performed to cool down the post cascade nuclear system by
  emission of protons,
  neutrons, and alpha particles, in competition with the fission process~\cite{Dep04}.
Briefly, the FSI were calculated through a Monte Carlo simulation of all
nuclear processes after the primordial NMWD~\cite{Dep02,Dep04,Gon97,de98, Dep01},
including important features, such as the Pauli blocking within both
elastic and inelastic nucleon-nucleon collisions~\cite{Dep02,Dep01}.

Note that the interplay  between the IPSM  and the
intranuclear cascade calculation is done  through  the initial configuration
for the Monte Carlo simulation, since the two intruder nucleons in initial sample
configurations have their  energies, momenta, and  opening angle
determined stochastically from  Eq. \rf{13}.
Another important bridge between the two models is the choice of the
isospins for the two nucleons that start the cascade process. This  is
also done via a Monte Carlo  sampling of the outgoing channel according
to the IPSM value for the $\Gamma_n/\Gamma_p$  ratio in the primary NMWD process.
\section{Result and discussion}
\begin{center}
\begin{figure}[htpb]
\begin{tabular}{cc}
\includegraphics[scale=0.4]{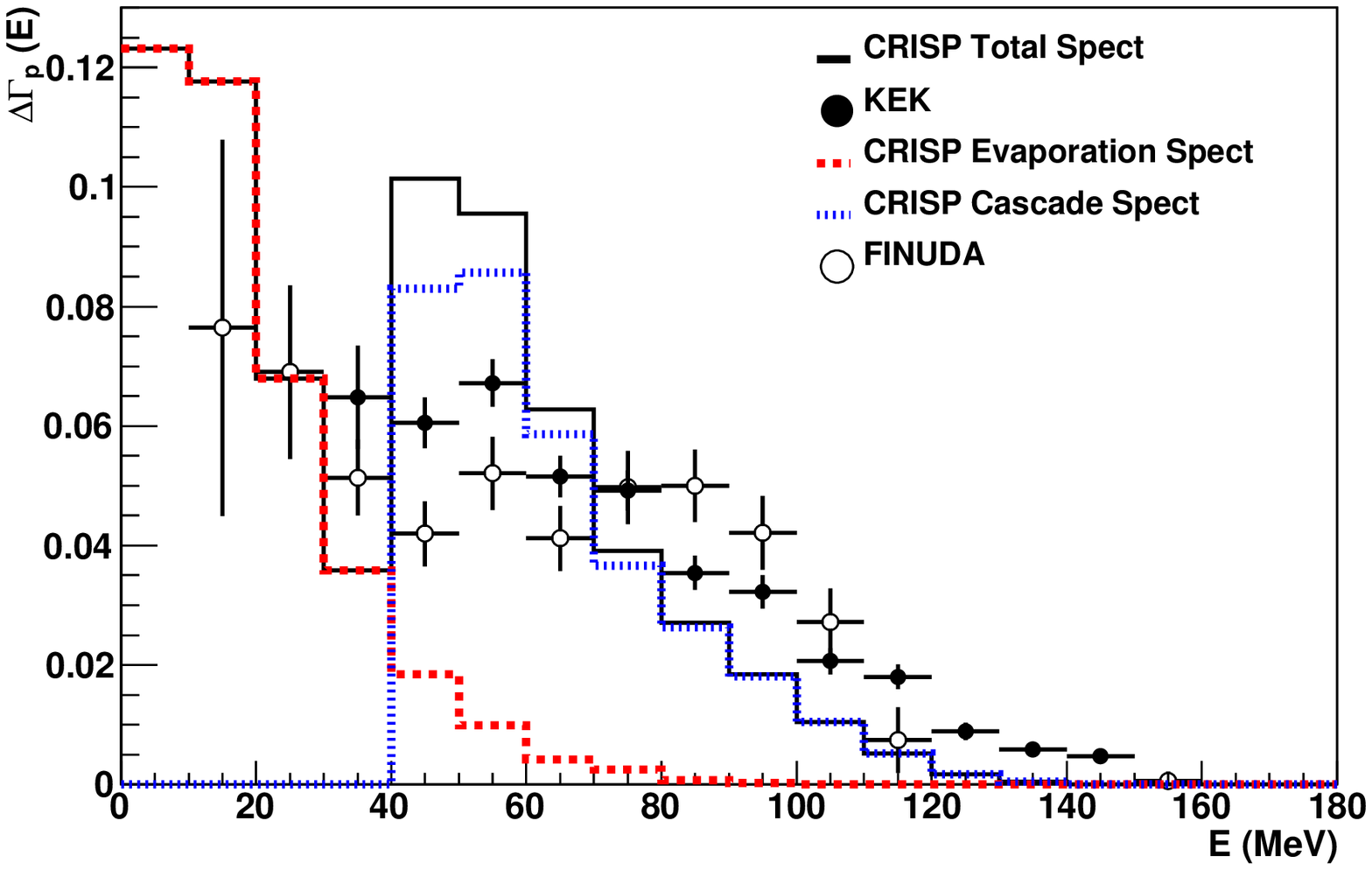}&
\includegraphics[scale=0.4]{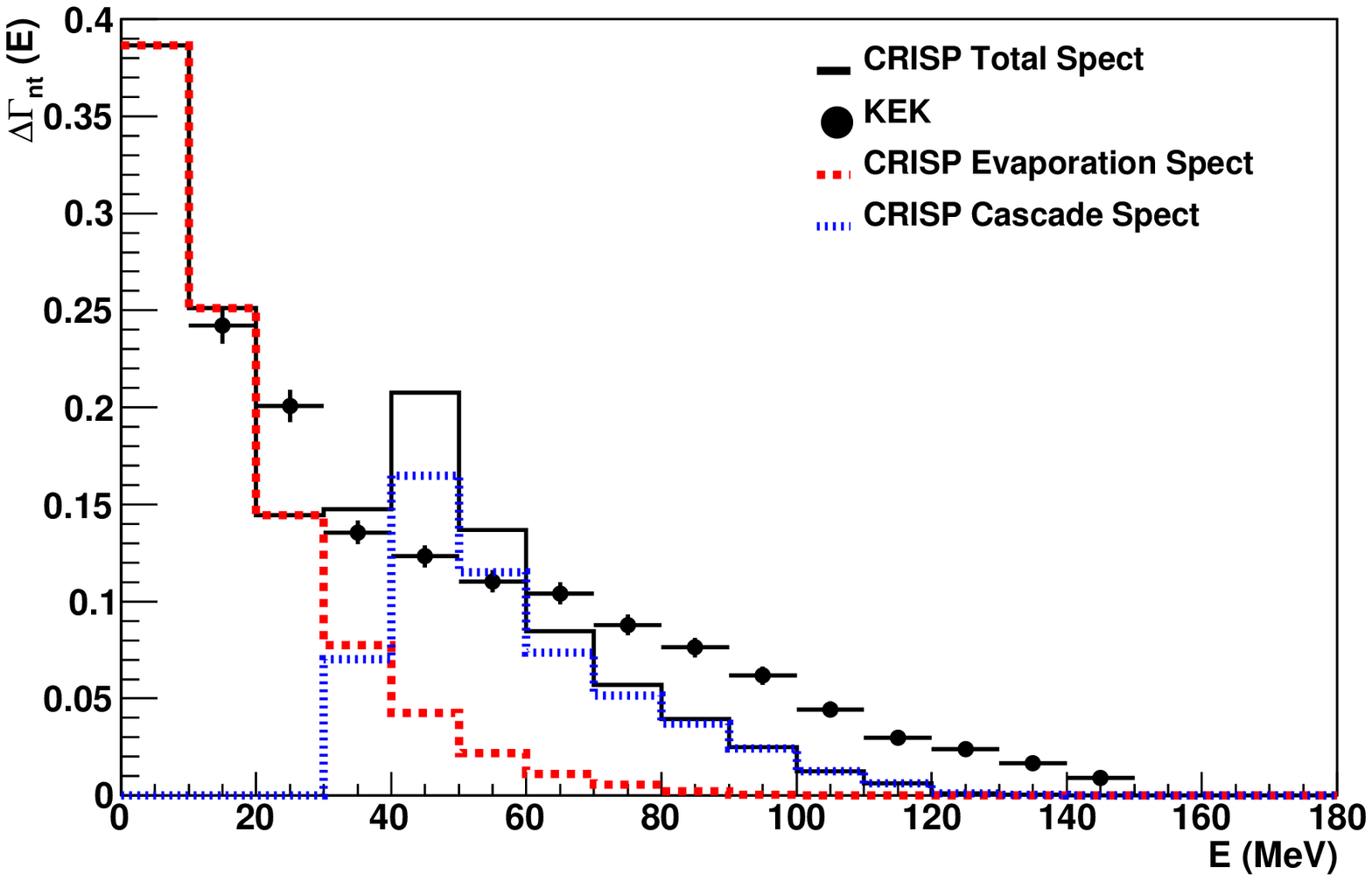}
\end{tabular}
\caption{Calculated inclusive single particle spectra (full histogram),
 compared to the data from  Ref.~\cite{Kim09,Bha11} in open circles,
 and from Ref.~\cite{Agn10,Gar10} in full circles.
 The dotted histogram are the contribution coming from cascade regime of
 the process and the dashed one depicts the contribution of particles
 evaporation of the warm post-cascade residual nucleus.} \label{Fig2}
\end{figure}
\end{center}
In Fig. \ref{Fig2} are compared the calculated inclusive
proton (left panel), and neutron (right panel) spectra with
the  KEK~\cite{Kim09,Bha11}, and FINUDA~\cite{Agn10,Gar10} data.
It is well known that the SFI modify the corresponding IPSM spectra
$\Delta\Gamma_N(E_{N})
=S_N(E_N)\Delta E$, with $S_N(E_{N})$ given by \rf{6}, by
 promoting particles from the high energy region towards the low energy region.
From the figure one can see that the FSI mechanism is engendered  by
the interplay between  contributions coming from
the cascade regime emission, and
 those induced by the evaporation of the warm residual post-cascade nucleus.
The enhancement at low energy induced by the first effect
agrees with data, while the quenching at high energy
produced by the second effect is too strong,  in disagreement with  the data.
We  also note that
the threshold for the opening of the cascade effect
is sharper in the $pn$ that in the $nn$ case due to the   Coulomb barrier.

\begin{center}
\begin{figure}[htpb]
\begin{tabular}{cc}
\includegraphics[scale=0.4]{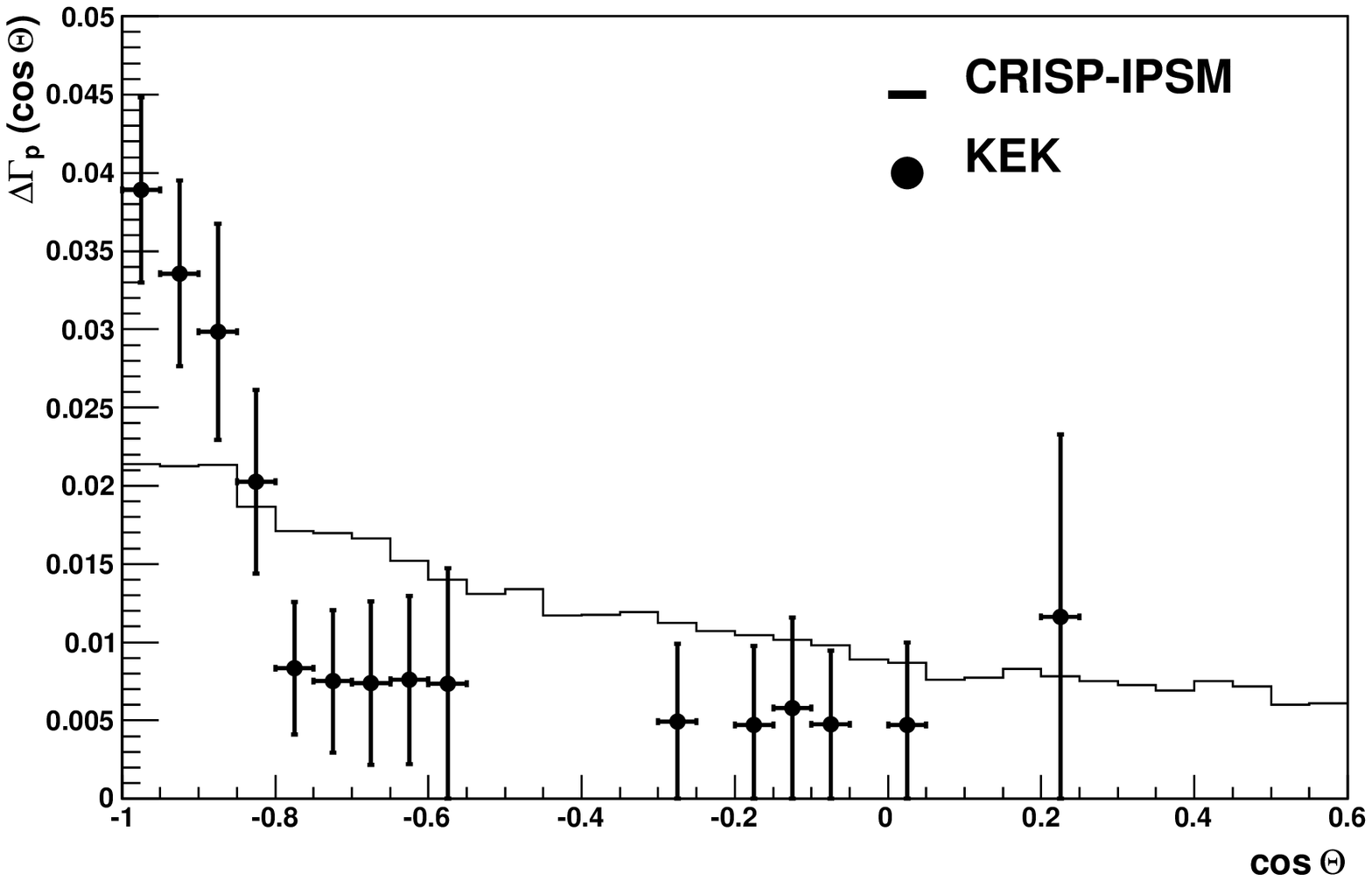}&
\includegraphics[scale=0.4]{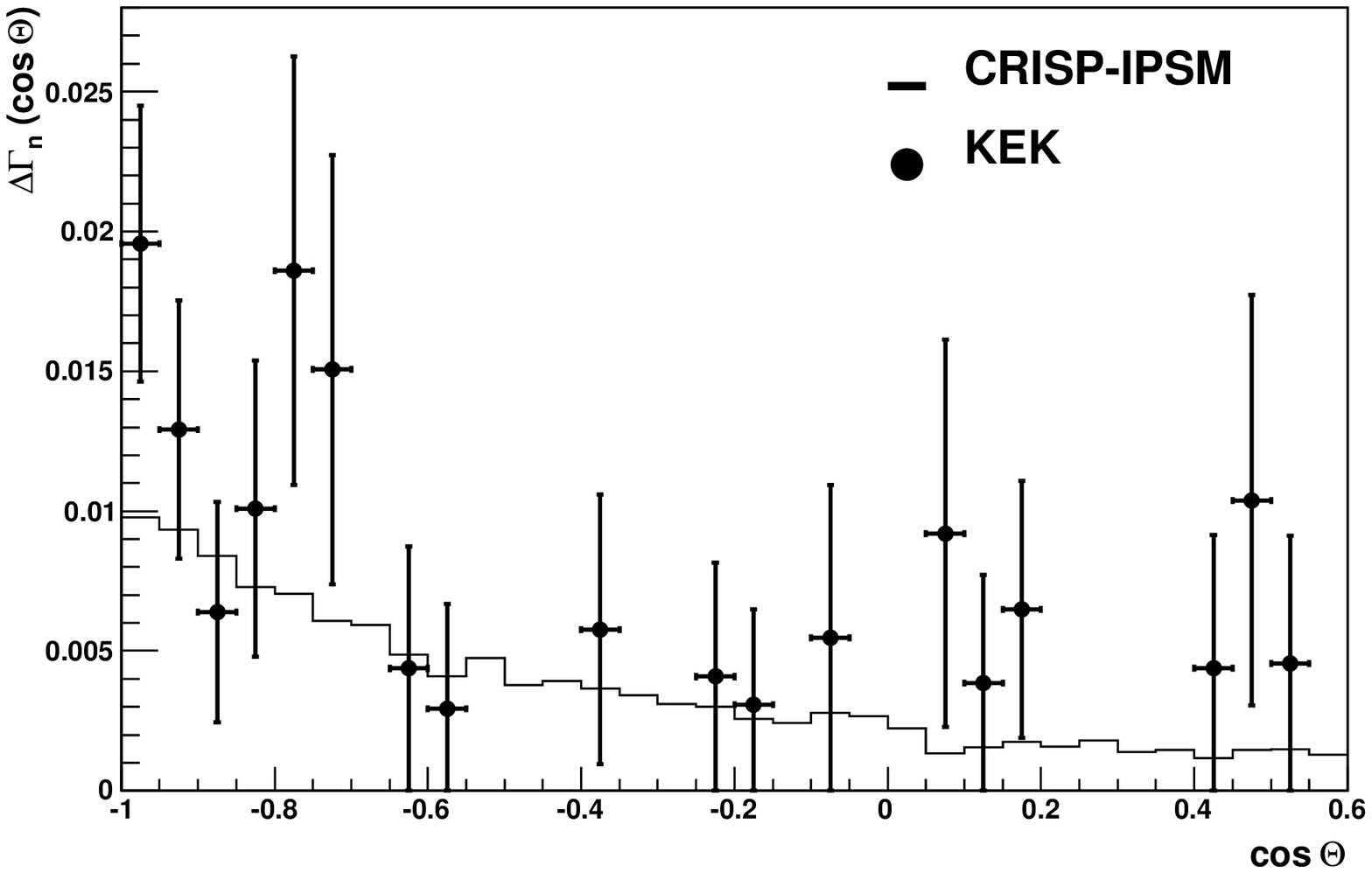}
\end{tabular}
\caption{Angular distribution of the  {np-pairs} (left
panel), and the {nn-pair} (right panel) compared to
experimental data from  Ref.~\cite{Kim09,Bha11}.} \label{Fig3}
\end{figure}
\end{center}

In left and right panels of Fig. \ref{Fig3}
 we show  the angular distributions for
the $nn$ and $np$ channels, respectively.
We get reasonable agreement with experimental results in the case of the $nn$ channel,
but  our calculation does not reproduce the peak observed
experimentally  in the back-to-back region for the $np$ channel.
This is due to too strong FSI quenching
on the IPSM spectra  $S_p(\cos\theta_{np})$ given by \rf{7}.

\begin{center}
\begin{figure}[htpb]
\begin{tabular}{cc}
\includegraphics[scale=0.4]{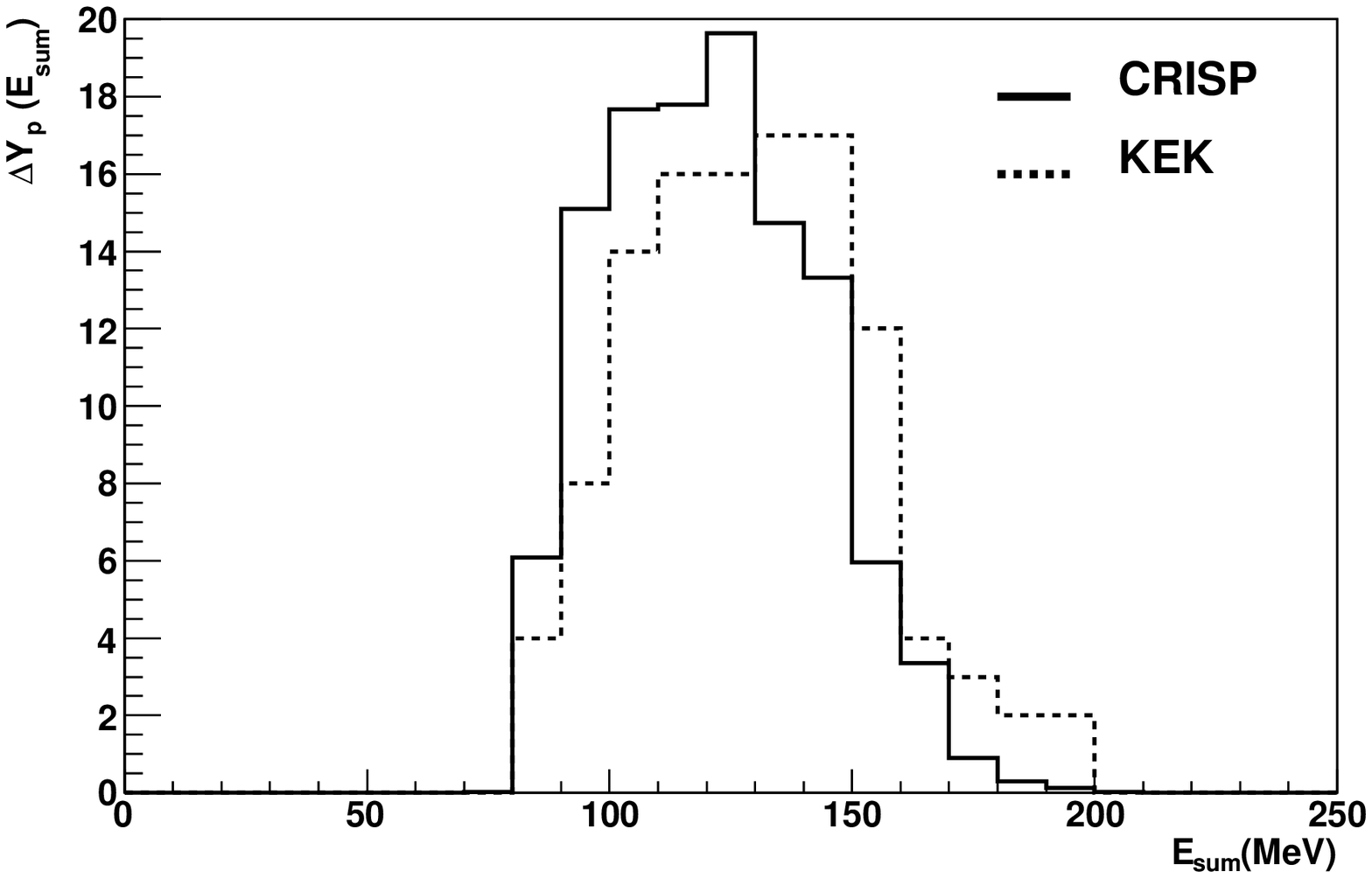}&
\includegraphics[scale=0.4]{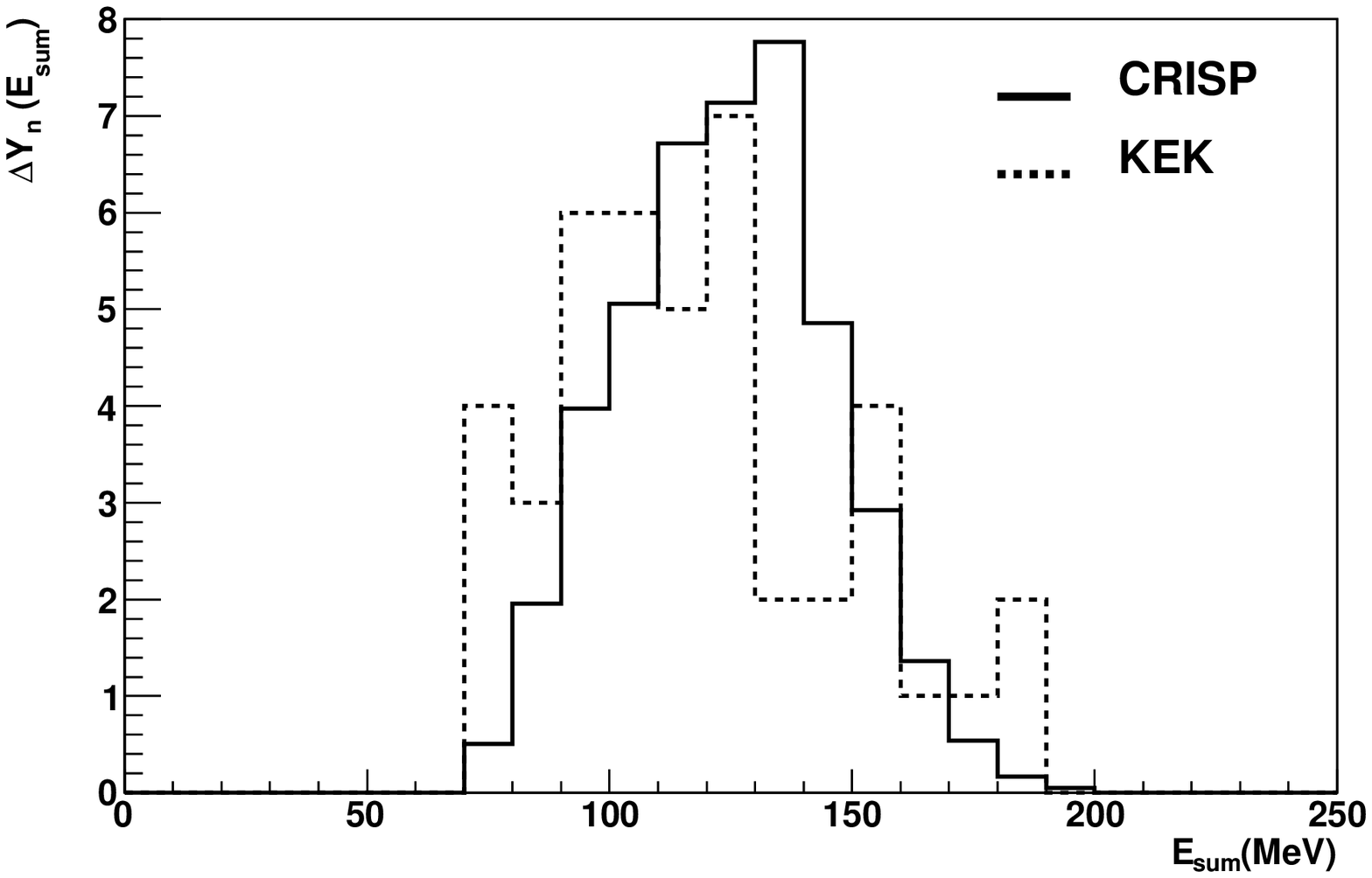}
\end{tabular}
\caption{The kinetic energy  distribution of the $nN$ pairs calculated for events
with strong back-to-back correlation of the opening angles. The full
histograms are the  results from the calculation, and dotted histograms are data
from Ref.~\cite{Kim06}. To define the back-to-back
events we have adopted the same  low energy
cutoff, and the same angle interval as done in the experiment.} \label{Fig4}
\end{figure}
\end{center}

In  the right panel of Fig. \ref{Fig4}
 we show the total kinetic energy of the correlated $np$ pairs,
  and compared them to the KEK experimental data~\cite{Kim06}.
  The theoretical predictions are normalized to the data according to Eqs. \rf{4}, and
  \rf{5}. We observed that both the centroid and width of the theoretical spectrum
  are quite similar to the experimental one, covering
   a broad energy interval from $\simeq 80$ MeV up to the $Q$-value at  $\simeq 150$ MeV.
  For the $nn$ channel, as  shown in the left panel of  Fig. \ref{Fig4}, the agreement
  between calculated and data is not so good as in the the $np$ case. However,
  we would like to draw attention to
  the fact that, because of very  low statistic, which in turn is due to
   difficulties  in detecting two neutrons in coincidence, the experimental histogram contains
   relatively small number of events.

We have also calculated the ratio between ${\Gamma}_{n}$ and ${\Gamma}_{p}$
by considering
 the emitted $nn$ and $np$ pairs in the back-to-back opening angle region
$\cos\theta_{nN} \leq -0.7$, getting the value $(\Gamma_n/\Gamma_p)^{\mbox{\tiny FSI}}
= 0.454 $,
 in nice agreement with the recently
reported experimental result $(\Gamma_n/\Gamma_p)^{\rm exp} = 0.53
   \pm 0.13$~\cite{Kim06},
  where the background uncertainty is $0.05$.
  At this point it is interesting to mention
  that  the IPSM model calculation yields
  the value $(\Gamma_n / \Gamma_p)^{\mbox{\tiny IPSM}} = 0.262$.
  We can attribute the difference between $(\Gamma_n / \Gamma_p)^{\mbox{\tiny IPSM}}$,
  and $(\Gamma_n / \Gamma_p)^{\mbox{\tiny FSI}}$
   to a possible depletion in the proton
  emission due the Coulomb barrier at the residual nucleus surface, implying
  that a neutron  produced during the primary process have higher probability of
  leaving  the nucleus than a primary proton.

\section{Conclusions}
The role played by the FSI on the  NMWD  has been investigated.
 For the inclusive nucleon spectra we have depicted  separately the
 contributions coming from  the cascade phase,  and the low energy ones  coming from
  the particle evaporation process in the post cascade phase regime.
 The importance of the latter nucleons in  reproducing
 the experimental data was put in evidence. In the analysis of angular distributions, we
 have obtained a reasonable agreement for the $nn$ channel, while  the results
 for the np channel still need to be improved.
  From the comparison of the  $\Gamma_n/\Gamma_p$ ratios before and after
  the inclusion of FSI,
 one sees that  these interactions increase the $n/p$ ratio by a factor of $\sim1.8$,
yielding  in this way the agreement with the measurement.

%

\end{document}